# China's SCI-indexed publications: facts, feelings, and future directions

Weishu Liu

School of Information Management and Engineering, Zhejiang University of Finance and Economics, Hangzhou 310018, Zhejiang, China

Email: wsliu08@163.com

**Abstract**

Purpose: In relation to the boom in China's SCI-indexed publications, this opinion piece examines this phenomenon and looks at future possible directions for the reform of China's research evaluation processes.

Design/Approach/Methods: This opinion piece uses bibliographic data for the past decade (2010–2019) from the Science Citation Index Expanded in the Web of Science Core Collection to examine the rise in China's SCI-indexed publications.

Findings: China has surpassed the United States and been the largest contributor of SCI publications since 2018. However, while the impact of China's SCI publications is rising, the scale of this impact still lags behind that of other major contributing countries. China's SCI publications are also overrepresented in some journals.

Originality/Value: Reporting the latest facts about China's SCI-indexed publications, this article will benefit the reform of China's research evaluation system.

Previous studies have shown a rapid rise in China's SCI- (SCI refers to Science Citation Index Expanded for brevity) indexed publications in recent years (Liu et al. 2015). Most recently, China's Ministry of Education and Ministry of Science and Technology have jointly issued a document to stop the practice of SCI-based evaluation in scientific research. This document has stimulated a wide discussion in academic circles around the world (Mallapaty 2020; Nature Editorial 2020; Zhu 2020). Some people are in favor of the decision; others, however, are skeptical. Supporters argue that it will switch the research evaluation practice from counting the number of SCI papers and citations to evaluating the research itself. Others, however, suspect that the science and education ministries are trying to destroy an objective indicator but without establishing a new and more appropriate one. Consensus on this issue, however, does not appear imminent. In this opinion piece, the author tries to uncover the facts, feelings, and future directions in relation to China's SCI-indexed publications.

## Three facts about China's SCI-indexed publications

Before any discussion about China's SCI-indexed publications, three recent developments should first be noted: China has been the largest contributor of SCI publications since 2018; while the impact of China's SCI publications is rising, it still lags behind that of other major contributing countries; and China's SCI publications are overrepresented in some journals.

### *China has been the largest contributor of SCI publications since 2018*

According to the Statistical Data of Chinese S&T Papers released annually by the Institute of Scientific and Technical Information of China (hereinafter referred to as the ISTIC report), China has been the second-largest contributor of SCI publications every year from 2009 to 2018. However, if we limit the document type to articles and reviews, which is a common practice in the scientometrics community (Bornmann & Bauer 2015; Zhang, Rousseau & Glänzel 2011), a different scenario emerges: China has surpassed the United States as the world's largest producer of SCI papers since 2018.

For comparison, this opinion piece also calculates the SCI publication outputs of the United States during the past decade (data accessed on March 5, 2020; only articles and reviews are considered). Figure 1 shows the annual SCI publication outputs of China and the United States. The rise in China's SCI publications is very noticeable: from 137,000 in 2010 to 482,000 in 2019, an increase of 252%. The data also shows that this rapid growth has actually accelerated in the last two years. In contrast, the output growth of the United States has been rather slower: from 304,000 in 2010 to 400,000 in 2019, an increase of 31%.

[INSERT FIGURE 1 HERE]

During the past 10 years, China has contributed 19.7% of the world's total SCI publications, while one quarter (25.1%) was contributed by the United States. According to Figure 2, China's share of global SCI publication output also rose rapidly, from 12.2% in 2010 to 27.5% in 2019, while the United States' share has gradually decreased, falling from 27.2% in 2010 to 22.8% in 2019.

[INSERT FIGURE 2 HERE]

*The impact of China's SCI publications is rising but still lags behind other countries*
Despite the significant growth in China's SCI publications, it would be misleading to conclude that China has replaced the United States as the world's leading scientific power. Quality matters more than quantity and the issue of publication impact remains paramount. This publication impact is usually measured by the number of citations. According to the ISTIC report, the average number of citations of China's SCI publications during 2009–2019 is 10.92, which is 9.2% higher than the data based on the period of 2008-2018 reported in the previous year. However, the average number of citations of China's SCI publications is lower than the world average (12.68) and China ranks only 16[th] among the 22 main producers of SCI publications. To sum up, the citation impact of China's SCI publications is rising but still lags behind the competition.

*China's SCI publications are overrepresented in some journals*
As demonstrated in Table 1, 6.67% of China's SCI publications during the past decade were published in the top 10 journals. The table also shows China's relative share of each journal's output during that decade. According to the data, 35.1% of the top 10 journals' publications were contributed by authors from China, which is 78% higher than China's average contribution to SCI publications globally (19.7%).

[INSERT TABLE 1 HERE]

However, China's relative shares among the top 10 journals vary significantly. The largest publishing outlet, PLoS One, has published 1.2% of all China's SCI publications, though China has only contributed 16.2% of all the publications in this open access (OA) mega-journal. However, China's contributions to each of the other nine journals exceed the benchmark value of 19.7%. It is worth noting that 91% of the publications in the OA journal *The International Journal of Clinical and Experimental Medicine*, published by e-Century Publishing Corporation, are contributed by China, while 65% of the articles published in *IEEE Access*, another OA journal produced by IEEE Inc., are also contributed by authors from China. The highest percentage contribution rate by Chinese authors is to *Acta Physica Sinica* (99.9%) but this is understandable since it is a local journal published by the Chinese Physical Society.

**Some feelings about China's SCI-indexed publications boom**
Introduced in the 1980s, the SCI-based evaluation system has enhanced the quality of basic science research in China (Gong & Qu, 2010). With the continuous investment of resources, China's scientific research capabilities have improved significantly over the past 30 years. At the same time, the SCI database has also significantly updated its content, including the introduction of more and more regional and OA journals. The limitations of SCI-related metrics still exist and have been widely discussed (Hu et al. 2018; Lariviere et al. 2016; Liu, Hu, & Gu 2016; Liu et al. 2018a, 2018b). As Goodhart's law states, 'When a measure becomes a target, it ceases to be a good measure.' Essentially, SCI-related metrics are still very useful but need to be used wisely. The famous Leiden Manifesto for research metrics also suggests that 'quantitative evaluation should support qualitative, expert assessment' (Hicks et al. 2015).

In China, institutions at varying stages of development adopt different research evaluation strategies. Institutions weak in scientific research may judge the research outputs solely based on whether they are SCI indexed. However, many institutions of medium scientific strength have adopted other indicators, such as journal impact factor quartile and number of citations, to assist the evaluation process of research output, while leading institutions tend to establish their own lists of recognized journals. Regardless of these different approaches, almost all institutions also use peer review to evaluate the research outputs, especially when promotion is being considered.

In China, many professions with strong practical attributes, such as clinicians, need to publish articles in order to meet the evaluation criteria. Young scholars in many middle- and lower-tier universities also need to publish as many papers as possible to improve their employment assessment and relieve economic pressure, while many graduate students need to publish SCI papers to meet graduation requirements. All this has contributed to the boom in China's SCI publication during the past decade. However, the issue of quantity vs quality remains and the boom in output has been accompanied by increasing problems in relation to research integrity in China (Lei & Zhang 2018; Tang 2019).

Publication output has also been boosted by some mega-journals, as shown in Table 1. The fact that more than 60% (and even 90%) of some journals' publications are contributed by authors from China is frankly astonishing. However, many of these journals charge high article processing fees, indicating the existence of possible conflicts of interest for both publishers and authors. In particular, the possibility that some publishers are being driven by financial reasons into lowering their acceptance standards needs to be considered.

**Future directions**

Evaluation is a difficult task and different institutions need to explore their own evaluation options in practice. Three directions may be considered in future evaluation practice.

First, various forms of contribution should be recognized, not just publication outputs. Different positions entail different responsibilities and therefore produce different forms of output and all contributions, in whatever form, should be recognized. In important cases, such as university promotions, stipulating SCI papers or national projects as an essential condition may not be appropriate.

Second, SCI-based metrics should be used wisely rather than abandoned. The Web of Science's citation indexes are powerful and invaluable; however, the SCI-based metrics should be used carefully in evaluation practice, with more focus on quality rather than quantity. Some SCI-based metrics, such as journal impact factor quartile and number of citations excluding self-citations, can act as important supporting indicators in the evaluation process. In practice, a blacklist or a grey list are effective ways of weeding out Web of Science indexed papers and journals of questionable quality.

Third, the peer review process needs to be optimized by combining quantitative and qualitative methods. While peer review is always an option for research evaluation, it is costly, time-consuming, and subjective. However, we should note that papers published in many prestigious journals have passed rigorous peer review. Evaluation management departments therefore need to pick appropriate and trusted experts and provide referees with both quantitative and qualitative information.

**Declaration of conflicting interests**
The author declares no potential conflicts of interest with respect to the research, authorship, and/or publication of this article.

**Funding**
This research was supported by the Zhejiang Province Natural Science Foundation (grant number #LQ18G030010).

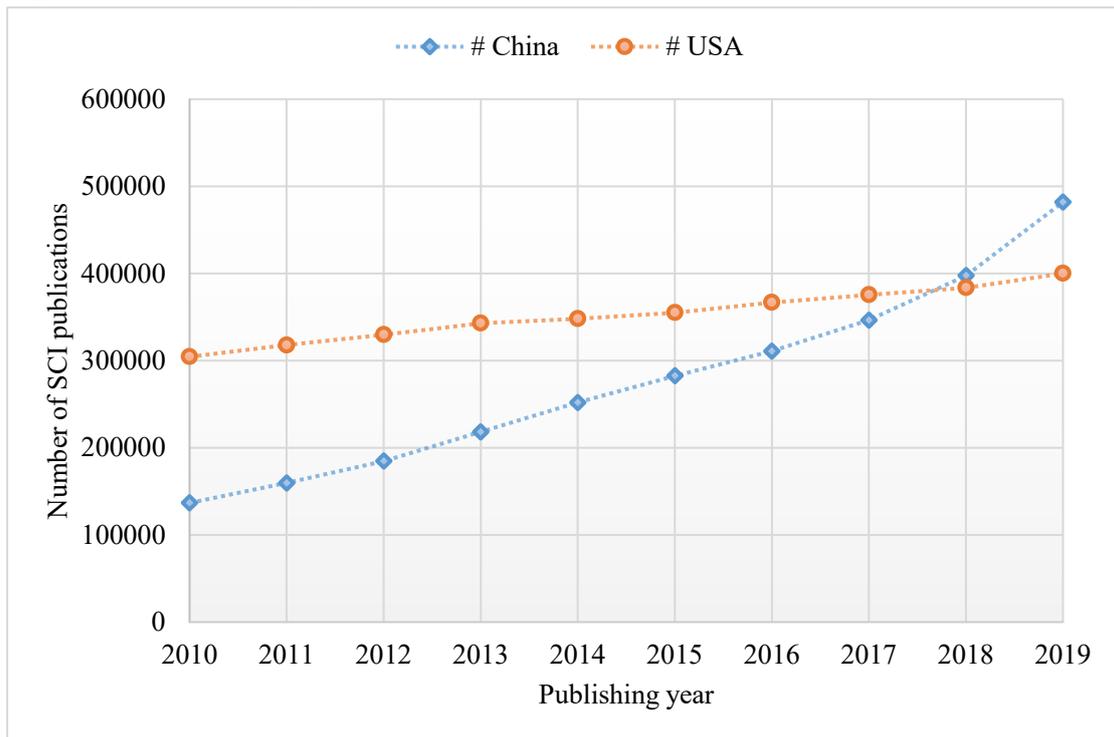

Figure 1 Numbers of SCI publications: China vs the United States

Note: Only articles and reviews are taken into account.

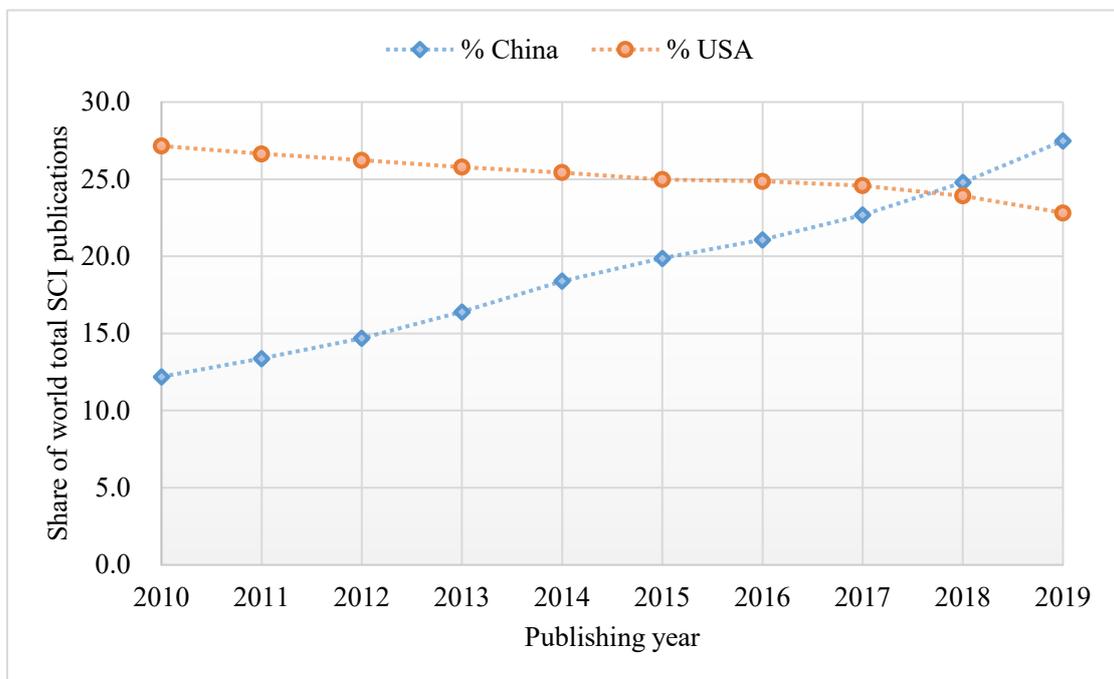

Figure 2 Relative shares of SCI publications: China vs the United States

Note Only articles and reviews are taken into account.

Table 1 Top 10 publishing outlets of China's SCI publications (2010-2019)

| Ranking | Journal | Record Count | % of China's total SCI pubs | Total pubs in each journal | China's share (%) | Publisher |
|---|---|---|---|---|---|---|
| 1 | PLoS One | 33274 | 1.20 | 204900 | 16.2 | Public Library Science, USA |
| 2 | RSC Advances | 28991 | 1.05 | 55582 | 52.2 | Royal Soc Chemistry, England |
| 3 | Scientific Reports | 28783 | 1.04 | 100537 | 28.6 | Nature Publishing Group, England |
| 4 | IEEE Access | 16433 | 0.59 | 25265 | 65.0 | IEEE Inc, USA |
| 5 | Journal of Alloys and Compounds | 15673 | 0.57 | 31246 | 50.2 | Elsevier, Switzerland |
| 6 | ACS Applied Materials Interfaces | 13843 | 0.50 | 29019 | 47.7 | Amer Chemical Soc, USA |
| 7 | Acta Physica Sinica | 12765 | 0.46 | 12778 | 99.9 | Chinese Physical Soc, China |
| 8 | Chemical Communications | 11990 | 0.43 | 31415 | 38.2 | Royal Soc Chemistry, England |
| 9 | Applied Surface Science | 11799 | 0.43 | 24356 | 48.4 | Elsevier, Netherlands |
| 10 | International Journal of Clinical and Experimental Medicine | 11385 | 0.41 | 12511 | 91.0 | e-Century Publishing Corp, USA |
|  | Top 10 | 184936 | 6.67 | 527609 | 35.1 | N/A |

Note Only articles and reviews are taken into account.